# High Throughput and Less Area AMP Architecture for Audio Signal Restoration

R.Swetha[1], D.Rukmani Devi[2]
[1]PG Scholar,[2]Professor, Department of ECE, RMK Engineering College
R.S.M Nagar, Gummidipoondi Taluk, Thiruvallur Dist., Kavaraipettai-601206.

***Abstract*—Audio restoration is effectively achieved by using low complexity algorithm called AMP. This algorithm has fast convergence and has lower computation intensity making it suitable for audio recovery problems. This paper focuses on restoring an audio signal by using VLSI architecture called AMP-M that implements AMP algorithm. This architecture employs MAC unit with fixed bit Wallace tree multiplier, FFT-MUX and various memory units (RAM) for audio restoration. VLSI and FPGA implementation results shows that reduced area, high throughput, low power is achieved making it suitable for real time audio recovery problems. Prominent examples are Magnetic Resonance Imaging (MRI), Radar and Wireless Communications.***

***Keywords*—Approximate message passing (AMP), field-programmable gate array (FPGA), very-large scale integration (VLSI)***

## I INTRODUCTION

The problem of identifying the sparse vector x $\in C^{Na}$ from M linear and non-adaptive measurements collected in the vector, as in [4],

$$z = Ax + Be \quad (1)$$

where A $\in C^{MXNa}$ and B $\in C^{MXNb}$ are deterministic and general (i.e., not necessarily of the same cardinality and possibly redundant or incomplete) dictionaries, and e $\in C^{Nb}$ is a sparse noise vector. The support set of e and the corresponding nonzero entries can be arbitrary; in particular, e may also depend on x and or the dictionary A.

### *A. Signal Separation*

The decomposition of audio or videosignals into two distinct components also fits into ourframework. Signal separationthen amounts to extracting the sparse vectors x and e,simultaneously, from z = Ax + Be, where z representsthe image to be decomposed, as in [4].

### *B. Narrowban Interference*

In many applications one isinterested in recovering audio, video, or communicationsignals that are corrupted by narrowband interference.Such impairments typically exhibit a sparse representationin the frequency domain, which amounts to settingB = $F_M$ in (1), where $F_M$ denotes the M XM discreteFourier transform matrix, as in [4].

Therefore our paper concentrates on recovery of information from sparsely corrupted signals with required throughput, power and area corresponding to the real time applications.

## II SPARSE SIGNAL RECOVERY ALGORITHMS

Sparse approximation also referred to as sparse decomposition is recovery of sparse signals from a system of linear equations or undetermined system which can be a given set of random matrices, as in [6]. Sparse approximation techniques have been used in wide range of applications such as image processing, audio processing, biology, and document analysis. Sparse Decomposition can be analysed in two ways 1) Noiseless Observations 2) Noisy Observations.

Different types of sparse signal recovery algorithms and its properties are analysed and explained below as in [7],

1)T-MSBL 2) EM-SBL 3)ExCov 4) CoSaMP 5)subspace pursuit 6) Approximate Message Passing(AMP) 7)Bayesian Compressive Sensing 8) Magic-L1 9)Hard Thresholding Pursuit(HTP) 10)Fast Bayesian Matching Pursuit (FBMP) 11)FOCUSS 12)Smooth L0Some of the algorithms needed to know some a priori information, and we fed these required algorithms with the required a priori information.

Details are given in the following list:

T-MSBL, EM-SBL, ExCov, AMP, and BCS: These algorithms do not require any priori information

CoSaMP, Subspace Pursuit: These algorithms are fed with the number of nonzero elements

Magic-L1: This algorithm requires SNR value to calculate the regularization parameter

FBMP: This algorithm requires true SNR value and non zero elements

FOCUSS: This algorithm is fed with only true SNR value

HTP, Smooth L0: This algorithm is applicable only for noiseless case and it fails for noisy case.

Our paper concentrates on AMP algorithm for audio signal restoration.

## III. COMPRESSED SENSING

Compressed sensing is a new field of research that has been used in many applications such as signal and image processing, as in [1]. The key feature of compressed sensing is to find the sparsest solution of an underdetermined system of linear equations.

Compressed sensing has gained attention in research community by enabling the sampling of sparse signals using fewer measurements then tan the Nyquist rate suggests, as in [6]. Moreover, compressed sensing lowers the cost of sampling and is used in wide real time applications. The main drawback is that it results in high computational complexity for recovery of sparse signals. Apart from compressed sensing many sparse signal recovery algorithms with low computational effort has been proposed.

## IV. SPARSITY SIGNAL RESTORATION

The principles of CS and sparse signal recovery can be used to recover signals that are corrupted by, e.g., impulse noise, clicks/pops, or saturation artifacts. Consider a corrupted signal z $\varepsilon R^M$, which can be modelled as referred in equation 2, as in [5],

$$z = Aa + Bb + n \qquad (2)$$

Here, the matrix A $\varepsilon R^{MXN}$ is used to sparsify the signal s = Aa to be restored, the matrix B $\varepsilon$ $R^{MXN}$ is used to sparsify the (unwanted) corruptions (e.g., clicks/pops), and n $\varepsilon$ $R^M$ is additive noise, which models small errors that cannot be sparsified in either basis, as in [5].

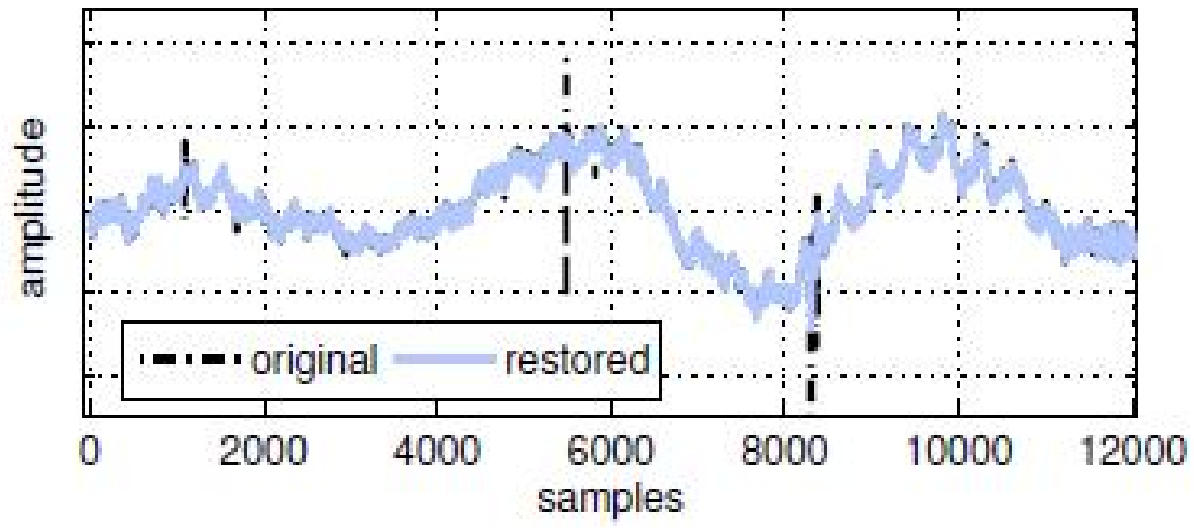

Fig.1 Performance and complexity of AMP for audio restoration, as in [5]

For example, a pair of matrices for audio signal restoration satisfying these properties corresponds to A being the MXM discrete cosine transform (DCT) matrix to sparsify audio and the corresponding restored signal is shown above Fig.1, as in [5].

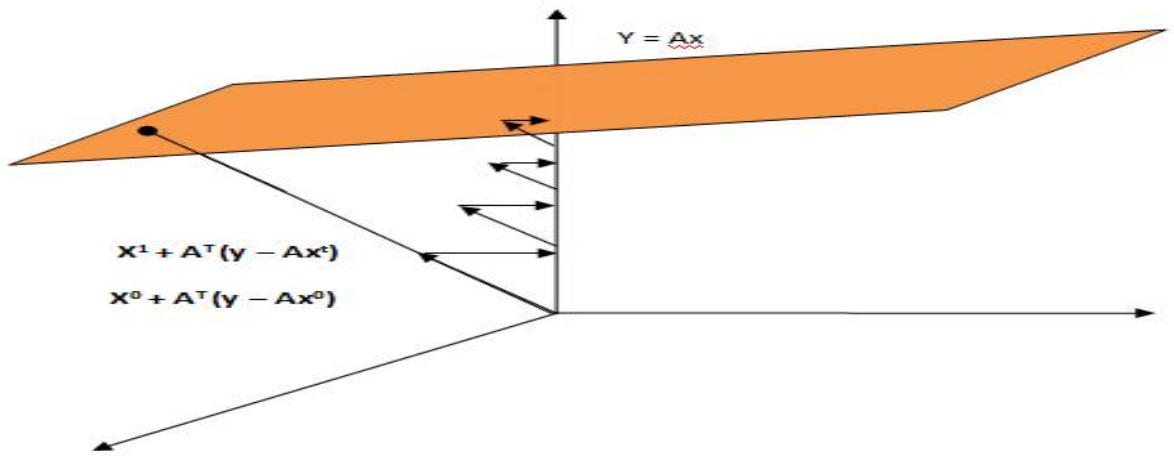

Fig.2 The pictorial representation for achieving sparest solution for iterative thresholding algorithms, as in [1]

For the restoration to succeed, the matrices A and B must satisfy the following conditions, as in [5]

i) Sparsify the signal s and corruptions e
ii) It should be incoherent

## V. APPROXIMATE MESSAGE PASSING ALGORITHM (AMP)

These are iterative algorithms, whose basic variables (“messages”) are associated to directed edges in a graph that encodes the structure of the statistical model. The relevant graph here is a complete bipartite graph over N nodes on one side (variable nodes), and n on the others (measurement nodes),as in [2].

MP has one important drawback compared to iterative thresholding. Instead of updating N estimates, at each iteration we need to update Nn messages, thus increasing significantly the algorithm complexity, as in [2].

This algorithm is called Approximate Message Passing or AMP. In the vector notation of this paper the algorithm can be written as in [1],

$$x^{t+1} = \eta(x^t + Az^t; \lambda^t), \qquad (3)$$

$$z^t = y - Ax^t + |I^{t-1}|/n\,(z^{t-1})$$

For the threshold parameter we can use one of the thresholding policies. As it can be seen this algorithm is very similar to the iterative soft thresholding algorithm. The only difference is in the extra term that is added to the calculation of $z^t$. This extra term although it barely adds to the computational complexity of our algorithm, has a significant effect on the phase transition of the algorithm significantly, as in [1].

## VI. STRUCTURE OF AMP ALGORITHM

We consider the model represented in Equation 4, as in [3],

$$y=As_o+w_o,\ s_o \in R^N,\ y,\ w_o \in R^n, \ (4)$$

With $s_0$ a vector that is ‘compressible’ and $w_0$ a noise vector.

This algorithm is interesting for its low complexity: its implementation is dominated at each step by the cost of applying A and A* to appropriate vectors. In some important settings, matrices A of interest can be applied to a vector implicitly by a pipeline of operators requiring N log (N) flops. Similar algorithms without this term are common in the literature of $s_0$ called iterative thresholding algorithms, as in [3].

The message passing term completely changes the statistical properties of the reconstruction, and it also makes the algorithm amenable to analysis by a technique we call State Evolution. Such analysis shows that the algorithm converges rapidly, much more rapidly than any known result for the IST algorithm, as in [3].

## VII. AMP-BASED ARCHITECTURES

There are two types of AMP architectures

i) AMP-M
ii) AMP-T

Both these architectures are optimized for an effective audio restoration and are suitable for arbitrary sparse signal recovery and are discussed below.

### *A. AMP-M*

This architecture implements multiply-accumulate units that is used to solve arbitrary sparsity based signal restoration and CS problems.

### *B. AMP-T*

It provides solution for cases where multiplications of a vector with D and $D^T$ that can be carried out using a fast transform.

But in our paper we have discussed about AMP-M architecture for effective audio restoration.

## VIII AMP-M ARCHITECTURE

Our paper concentrates on AMP-M architecture shown in Fig 3.

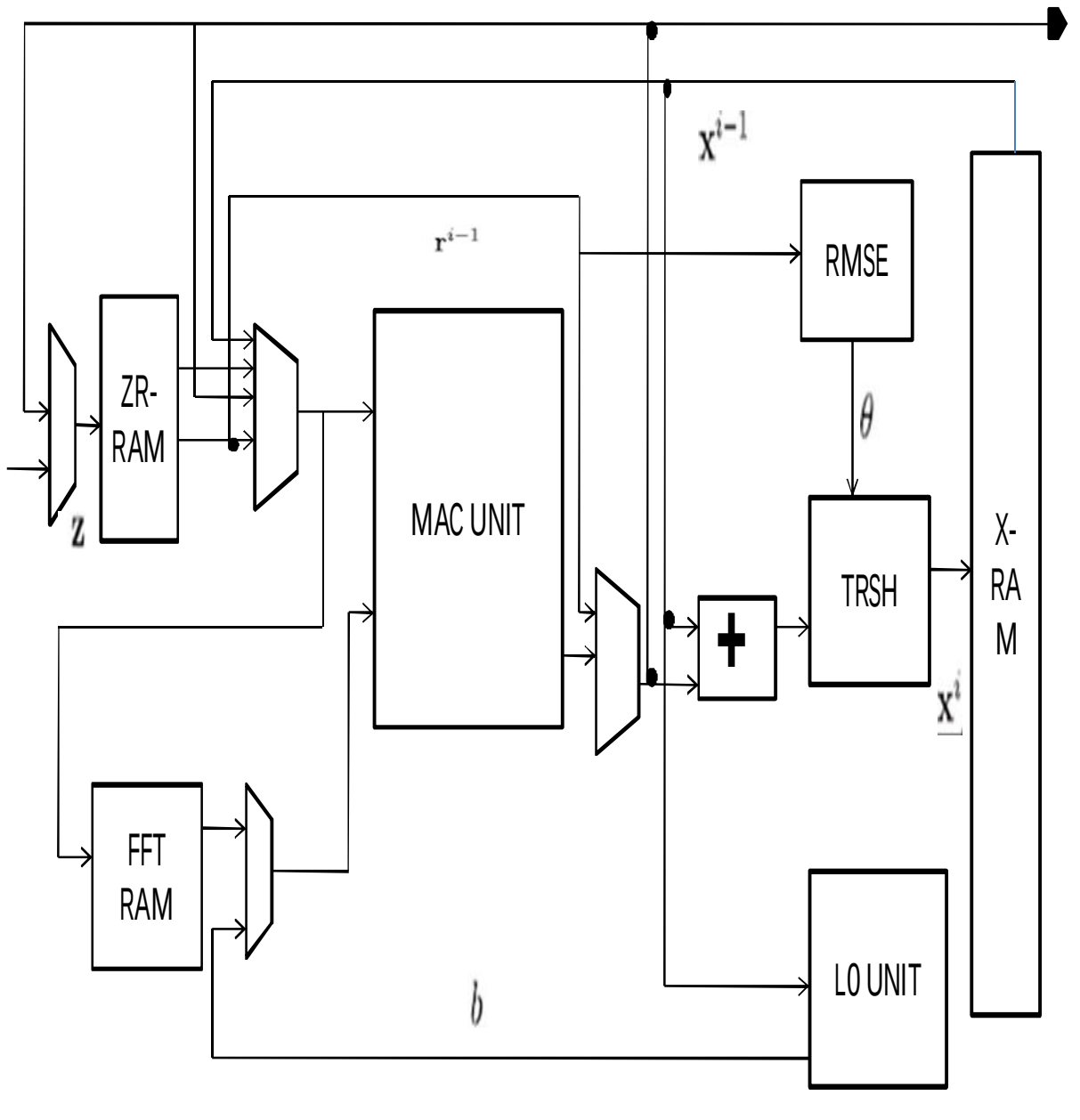

Fig.3 AMP-M Architecture

The above architecture implements AMP-M algorithm which has the following steps such as

1. Initialization
2. Determine the number of iterations
3. Calculation of residual value
4. Estimation of thresholding parameter
5. Determination of current signal estimate
6. Repeat the process until RMSE is less than or equals to ET parameter set to 0

ET parameter is called as Early Termination which reduces the complexity of AMP algorithm. AMP iterations can be terminated as soon as the RMSE is small enough. Therefore we can set threshold parameter $\gamma \geq 0$ and stop the iterative process as soon as RMSE $\leq 0$, as in [6].

This process continues until audio signal is restored which is the approximation of the input signal.

## IX. MODULES DESCRIPTION

The various modules used in AMP-M Architecture shown in Fig.3 are explained below in detail.

ZR-RAM: is used to store the input variables $r_i$ (residual value) and z.16-bit input variables are assigned to this architecture.

FFT-RAM: The input values from ZR-RAM are applied to this unit which generates a value corresponding to the input value by using M/2 FCT/IFCT algorithm. This algorithm does the following operations such as Reorder, Reduce, FFT/IFFT, expand and rotate to generate the appropriate output values. The real valued input vector are first reordered and stored to a vector c'. Then the reordered vector is converted into a complex-valued vector c half of the length. The main task of this algorithm is to compute an M/2 length FFT of the vector c. The result is then expanded into conjugate symmetric vector f'. To obtain the result the entries of f' are rotated by certain twiddle factors, as in [6].

MAC UNIT: MAC Unit is employed in various fields such as digital signal processing that finds its use in video and image processing. It is also used in VLSI architecture, that requires an required set of input signals to generate the appropriate outputs. The main components of MAC unit are Multiplier, adder and accumulator. The function is multiplication, addition and the following results will stored or fed to the accumulator which inputs the results to other modules or components.

We use Wallace tree multiplier which implements fixed 16 bit operation which reduces area, power and time complexity.

The main function is to multiply two integers. It comprises of 3 steps such as

i) Multiply

ii) Reduce

iii) Group

TRSH UNIT: TRSH unit is used to perform thresholding function. To implement this function, it uses subtract-compare-select unit that applies thresholding in a serial and element-wise manner.

RMSE (ROOT MEAN SQUARE ERROR):The main function of RMSE is to obtain the differences between the predicted value and the actual values observed.

X-RAM: X-RAM is a single port RAM that stores the next signal estimate and the result is fed as input to the Z-RAM.

## X. RESULTS AND DISCUSSION

### *A. SOFTWARE IMNPLEMENTATION USING MODEL SIM*

Model Sim is the software tool used for simulation. The benefits of this tool are:

- High performance HDL simulation solution for FPGA & ASIC design teams
- The best mixed-language environment and performance in the industry

During memory write operation, when 'we' signal is asserted high (we=1), the input values Zin & Rin are written to the memory locations mem1 and mem2 respectively. The signal re=1 denotes memory read operation. By reading the values from the memory, it outputs Zout & Rout. Both 'wclk' and 'rclk' are set high initially for complete audio restoration.

The count value is incremented for every positive edge of the clock or reset signal. This incrementing process helps for signal recovery. From the Fig.4, it is evident that the architecture restores to original values at the output for the corresponding input values applied, which shows the improved performance of AMP based architecture.

The above process is repeated in a stepwise manner such that it provides an effective audio restoration. In order to restore a corrupted 16 bit stereo audio signal at 44.1 k sample/s in real time, we developed an efficient FPGA implementation of the AMP algorithm, as in [5].

Signal restoration is performed using the DCT–identity pair on (independent but overlapping) blocks of length M = 512, which was found to deliver good restoration quality at reasonable hardware costs, as in [5].The implementation performs a maximum of Imax = 28 AMP iterations, which ensures good convergence.

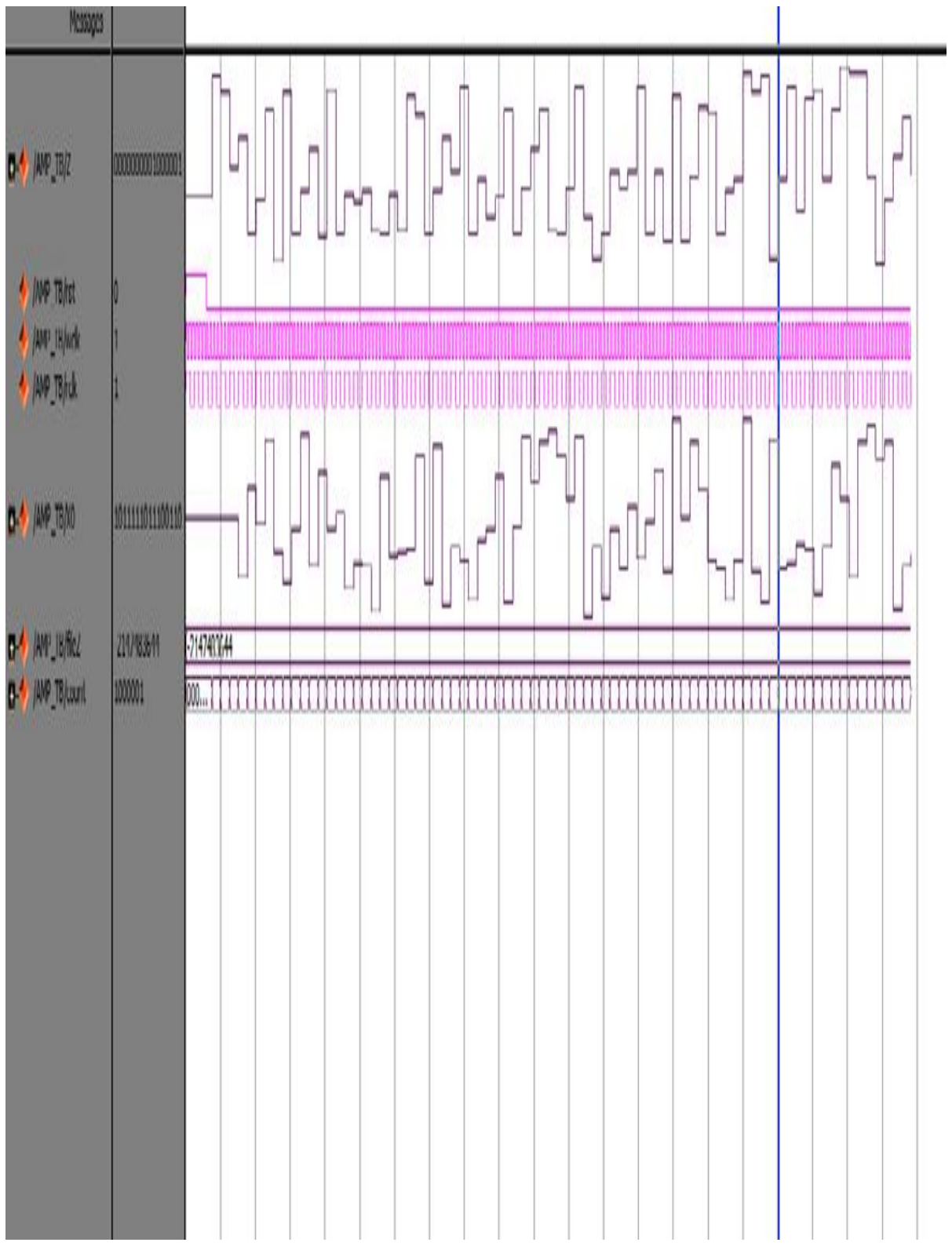

Fig.4 Simulation Results for the AMP-M architecture

The function of effective audio restoration using AMP algorithm according to specific criterion is shown above Fig. 4.

### *B. HARDWARE IMPLEMENTATION RESULTS USING ALTERA QUARTUS II*

The DE0 Board is used for development and educational purposes. It is an Altera Cyclone III 3C16 FPGA device. This board provides 15,408 logic elements, 346 I/O pins. Furthermore, it has various features making it as a sophisticated tool and is used in various research fields.

The hardware implementation results are analyzed and shown below.

*C. FINAL REPORTS*

The below Fig. 5 depict the total area required for restoring an audio signal using our proposed AMP-M architecture. Out of 15,408 logic elements present in the tool, it makes use of 3693 logic elements. Therefore the proposed Architecture consumes 24% of the total area, thereby satisfying less area requirements.

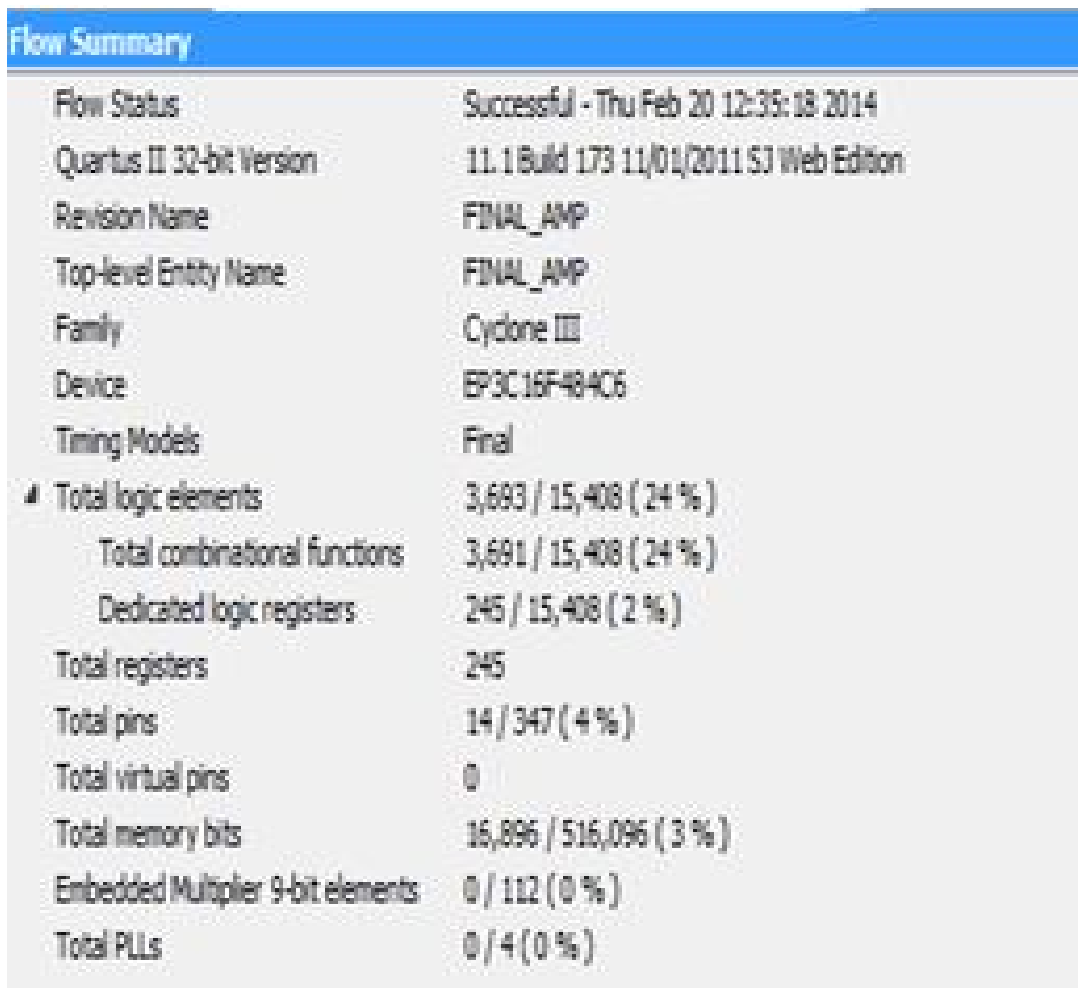

| Flow Summary | |
|---|---|
| Flow Status | Successful - Thu Feb 20 12:35:18 2014 |
| Quartus II 32-bit Version | 11.1 Build 173 11/01/2011 SJ Web Edition |
| Revision Name | FINAL_AMP |
| Top-level Entity Name | FINAL_AMP |
| Family | Cyclone III |
| Device | EP3C16F484C6 |
| Timing Models | Final |
| Total logic elements | 3,693 / 15,408 ( 24 % ) |
| Total combinational functions | 3,691 / 15,408 ( 24 % ) |
| Dedicated logic registers | 245 / 15,408 ( 2 % ) |
| Total registers | 245 |
| Total pins | 14 / 347 ( 4 % ) |
| Total virtual pins | 0 |
| Total memory bits | 16,896 / 516,096 ( 3 % ) |
| Embedded Multiplier 9-bit elements | 0 / 112 ( 0 % ) |
| Total PLLs | 0 / 4 ( 0 % ) |

Fig.5 Final Area Report

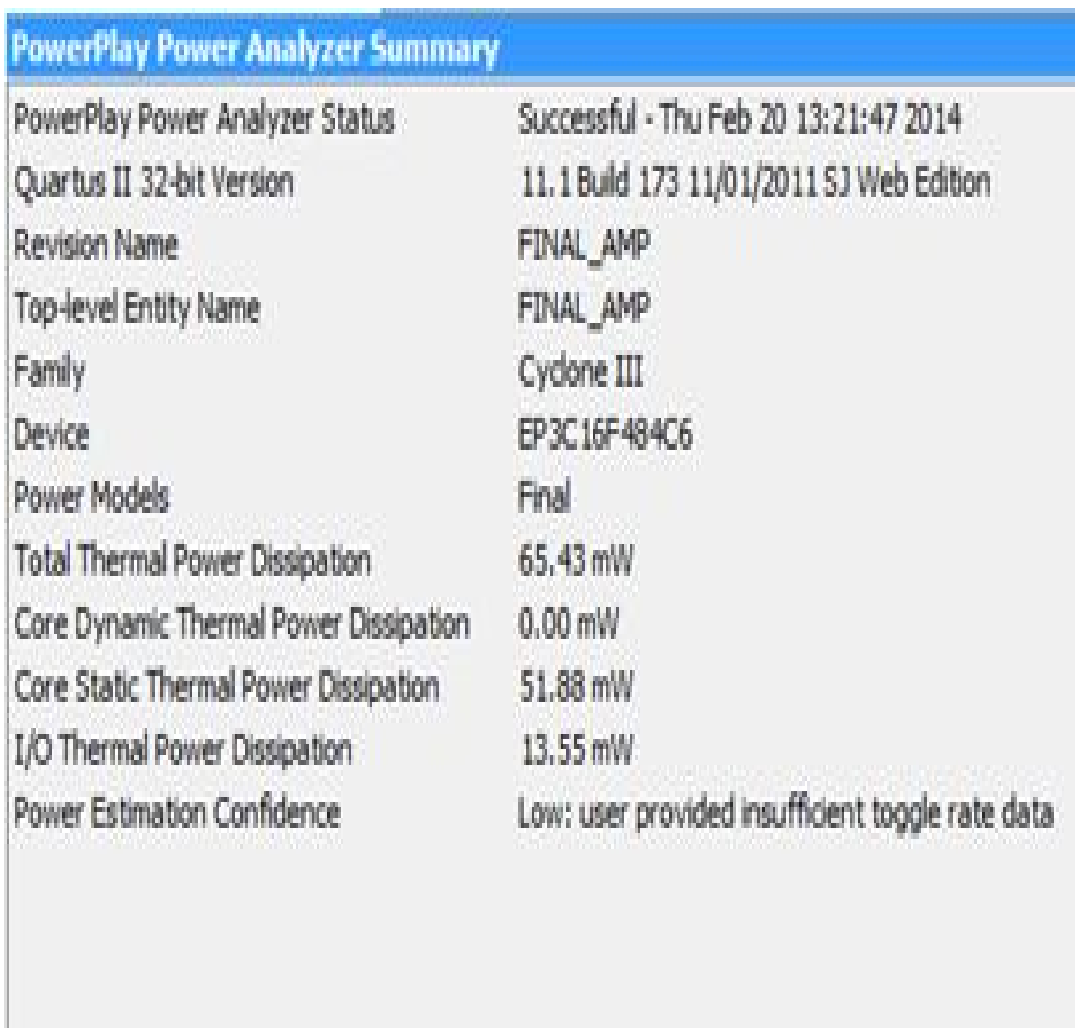

| PowerPlay Power Analyzer Summary | |
|---|---|
| PowerPlay Power Analyzer Status | Successful - Thu Feb 20 13:21:47 2014 |
| Quartus II 32-bit Version | 11.1 Build 173 11/01/2011 SJ Web Edition |
| Revision Name | FINAL_AMP |
| Top-level Entity Name | FINAL_AMP |
| Family | Cyclone III |
| Device | EP3C16F484C6 |
| Power Models | Final |
| Total Thermal Power Dissipation | 65.43 mW |
| Core Dynamic Thermal Power Dissipation | 0.00 mW |
| Core Static Thermal Power Dissipation | 51.88 mW |
| I/O Thermal Power Dissipation | 13.55 mW |
| Power Estimation Confidence | Low: user provided insufficient toggle rate data |

Fig. 6 Final Power Report

The belowFig. 6 shows the total power required to restore an audio signal using proposed AMP-M architecture. For proper audio restoration the power required is 65.43 Mw which is comparatively low compared to other techniques/architectures implemented.

The below Table I shows the compares the results of Quartus –II 32-BIT Version with the results implemented in IP8M 65nm CMOS technology. From the above table it is evident that reduced power of 65.43 Mw and a high throughput of 408.5 MHz is achieved, thus satisfying our requirements.

TABLE I

| REQUIREMENTS | QUARTUS –II 32-BIT VERSION | IP8M 65nm CMOS Technology |
|---|---|---|
| FINAL POWER [Mw] | 65.43 | 177.5 |
| FINAL THROUGHPUT [MHz] | 408.5 | 333 |

Table I Comparison Results of Quartus II with IP8M 65nm CMOS Technology

## XI. CONCLUSION

The AMP-M architecture is suitable for the recovery of signals acquired by compressive sensing or signal restoration problems relying on unstructured matrices. Optimized audio restoration using AMP-M based architecture has been presented in this project. Audio restoration can be achieved by using both software and hardware tools. The above process can be used in many real time applications. Therefore, by implementing the above AMP-M based architecture in Altera Quartus II (DE0) Board, required power, throughput, area, and clock frequency is achieved.